# Classical Economics: Lost and Found

## Comments

# Classical Economics:

## Lost and Found

Sabiou M. Inoua and Vernon L. Smith[1]

Chapman University

**The Indeterminacy of Neoclassical Value Theory**

At mid-twentieth century, the neoclassical mathematical theory of value culminated in Arrow and Debreu's (1954) model, which characterizes the static general equilibrium state of an economy. However, this description is unsatisfactory unless markets converge to such an equilibrium. That the Arrow-Debreu model could not accomplish this is an implication of the important result by Sonnenschein, Mantel, and Debreu, also known as the *SMD Theorem* (1972, 1973a, 1973b; 1974; 1974).[2] The heart of the problem is that the principle of individual utility maximization has no interesting implication for aggregate market demand, not even the law of demand; in fact, demand is essentially arbitrary in this theory. But this aggregation problem is often, if unintentionally, evaded through the artifice of the Representative Agent (Arrow, 1986;

---

[2] For reviews of the SMD Theorem, see Kirman (1989), Shafer and Sonnenschein (1993), and Rizvi (2006).

Kirman, 1992). Yet the problem is endemic, leading Frank Hahn (1982) to suggest that "we shall have to conclude that we still lack a satisfactory descriptive theory of the invisible hand" (746). Going further, F. M. Fisher (2013) concludes that "we do not have an adequate theory of value, and there is an important lacuna in the center of microeconomic theory. Yet economists generally behave as though this problem did not exist." (35)

We want to record the observation that acting as if a "problem does not exist" is not unique to economics or any science, for generally: "Scientists…do not abandon a theory merely because facts contradict it…they direct their attention to other problems." (Lakatos, 1978, 4)

We argue that neoclassical value theory suffers from a more basic and serious logical indeterminacy, which is inherent in the axiom of price-taking behavior, and which renders price dynamics indeterminate before inquiring as to its stability. If everyone in the economy takes price as given, whence come these prices? Who is giving these prices? Jevons avoided the indeterminacy by assuming that people must have complete information on supply and demand, and the consequent equilibrium prices—'perfect competition.' Walras in effect imported an external agent who found the prices by trial-and-error-correction (the Walrasian Auctioneer). Paradoxically, both approaches had the potential better to serve central planning, than a market economy. A theory based on price taking agents required some agency for giving prices. Indeed, the fit with socialism was rigorously established by influential neoclassical authors starting from Wieser (1893, ch. VI) and Pareto (1897, 364-371; 1909, 362-364), and, more formally during the Socialist Calculation Debate, by Barone ([1908] 1935), Lerner (1934), and Lange (1936, 1937). The paradox is hidden in the idea of 'perfect competition' a passive treatment of individuals who are not even interacting, let alone interacting in a rivalrous manner. 'Perfect competition' is the negation of any real competition, as Hayek (1948) emphasized.

**Experimental Market Economics**

While theoretical economics failed to provide a satisfactory account of the market mechanism, experimental economics, beginning mid-20th century established the remarkable stability, efficiency, and robustness of the market process under ordinary trading rules (notably the double auction institution, that existed apart from the economics literature) and under realistic market conditions (Smith, 1962; Plott, 1982; Smith, 1982; Davis and Holt, 1993).[3] These findings could not have been further from the expectations of the early neoclassic-trained experimentalists; their first struggle was with the failure to confirm widely shared expectations based on Jevons-Walras utility theory. Laboratory markets typically involve but a few buyers and few sellers, who know only their personal values and costs, or *reservation prices*, and who are obviously making the prices through their bids and asks. According to the neoclassical theory, we should expect only 'market failures' in this context. Yet the experimental markets generally converge to maximum efficiency. We argue that these experimental findings are in reality a corroboration of the classical view of the market mechanism (Inoua and Smith, 2019; Smith and Inoua, 2019). In fact wherever we examine an account of market price formation in the neoclassical literature itself, the authors invoke the old view; for example, Marshall [1890] 1920, bk. V, ch. 2) and the Austrian marginalists, all articulated explicitly a price discovery process through collective 'higgling and bargaining' in essentially the same way that the classical economists explained it. The modern introductory textbook, in describing equilibrium tendencies, also appeals to the outbidding and underselling of buyers and sellers in disequilibrium.

---

[3] The experiments—necessarily required to be very explicit about the instructional rules governing the exchange process—motivated new theoretical explorations; one recent example stemming directly from experimental findings is Anufriev, et al. (2011). Their "closed book, limited information" version of the theory, we suggest, most evidently to Adam Smith's ([1776] 1904, Chapter VII) informal conception of information exchange and response in a market. Their induced reservation price framework follows classical economics, as did the early experiments.

**Rediscovering Classical Economics**

The neoclassical school replaced the classical one; yet its root deficiencies are direct consequences of its departure from the older paradigm. First, concerning method. The classical economists, notably Adam Smith, observed carefully the market economy, and derived from the acute observations a theory of markets. Thus, Smith and his followers observe the experience of market participants, and discover the sophisticated, unintended, collective regularities that *emerge* from myopic, self-interested, individual behaviors and interactions, finding that it is the cause of specialization—the famous "invisible hand" metaphor, referring to agents' lack of awareness of what they have wrought. In place of this process, the neoclassical school in effect substituted thought experiments about an imaginary economy whose regularities rest entirely on artificially sophisticated individual rationality, and, epitomized by one idealized person (Robinson Crusoe, Walrasian Auctioneer, Representative Agent, Social Planner), and evoking a command economy.

**The Classical View on Demand and Supply**

Before the marginal revolution, the primitive concept in the theory of value was an individual's *valuation* of a good; i.e., *value in use*, which means the value that a person attaches to a good in view of the good's personal usefulness, as measured by the amount that this person is *willing to pay* (give up) in exchange for the good. Demand from Adam Smith to Jules Dupuit is given by *willingness to pay*. Malthus, for example, stated in his *Principles* that "demand will be represented and measured by the sacrifice in money which the demanders are willing and able to make in order to satisfy their wants" ([1820] 1836, 62). In his memoirs on utility, Dupuit (1844, 1849), while refining an intuition of J.B. Say, emphasized that use-value is given by maximum willingness to pay, namely by what we call today the reservation price. J.S. Mill reached the same conclusion but put it in a more technical way: "Value in use […] is the extreme limit of value in exchange", that is, price ([1848] 1909), bk. 3, ch. 1, § 2). Or: "the utility of a thing in the estimation of the purchaser, is the extreme limit of its exchange value" (bk. 3, ch. 2, § 1). Given the reservation price as a primitive concept, an individual's demand is very simple: by definition, one is willing to buy a good if one attaches to it a higher value than the price at which it can be acquired. Market demand is the distribution function of the demanders' reservation prices. Symmetrically, market supply is the distribution of the suppliers' reservation prices or costs. Experimental supply and demand functions are exactly of this nature. (Inoua and Smith, 2019; Smith and Inoua, 2019)

The demand side of this probabilistic view is particularly explicit in the works of the often neglected French contributors to the classical school[4], starting from Germain Garnier, who translated into French the *Wealth of Nation*s which inspired his *Abrégé élémentaire des Principes de l'économie politique* ([1796] 1846). In the second edition of this book (1846, 195-196), Garnier derived the law of demand from the distribution of willingness to pay, expressed as a fraction of wealth, which distribution he represented as a pyramid. Thus Garnier, following the classicalists, did not make the neoclassical error of treating consumption as an expenditure out of income. He and the classicalists understood the elementary proposition that a change in wealth is a consequence of spending, or not spending, out of total resources. Income, as change in wealth, is to be determined along with consumption by market actions.[5] The neoclassical fixation on utility maximization subject to a binding income constraint was the thought process that defined a static economy. For if the constraint is wealth, then the theory could not avoid actions causing change over time, unless wealth was also static, leaving theory without a means of understanding the market foundation of "the wealth of nations," yielding a complete negation of its classical purpose! In this sense, neoclassical economics was incoherent.

---

[4] This French tradition on demand, and value more generally, is thoroughly covered in Ekelund Jr and Hébert (1999), not as a part of the classical school, but as an anticipation or even the origin of the marginal-utility view on demand.

[5] The constraint language drifted into a "budget" constraint, but if you consume less than your budget, you are saving, and adding to wealth. Non-satiation assured that the budget was binding; again, an error for it afforded no value to saving. Hence, a static model minimally requires actors to have a willingness to pay for saving, whereupon not consuming is an active part of the opportunity set in the current period. In this way, the consumer choice problem is separated from its future, and left open to alternative ways of modeling why people save.

J.B. Say and Jules Dupuit also adopted explicitly this traditional view on demand (Say [1803] 2006), vol. II, bk. II, ch. 1; ([1828] 2010), vol. I, part III, ch. 4; Dupuit, 1844, 1849). Cournot's ([1838] 1897) treatment of demand also comes down to this view. Cournot, however, emerges as a pivotal figure in the transition from classical to neoclassical economics: his view on the source of demand is rigorously classical, though he went on to inspire much of the neoclassical theory of supply (Smith and Inoua, 2019). In a fascinating paragraph of his *Researches* ([1838] 1897), 49-50), he observed that market demand can be assumed to be a smoothly decreasing function of price (by the law of large numbers), even though individual demand is realistically discontinuous.[6]

---

[6] Significantly for our "lost and found" thesis, Cournot was the first to write the collection of individual reservation prices as a demand function, D = f (p). The classical economists, focused on the roots of action, and described the market process in terms of the experience of buyers and sellers who compare public prices with their reservation values and respond in bargaining. For that discovery process to play out, no one in a market needs to know that an economist/statistician can write D = f(p) as a representation of all market reservation prices. However, Cournot was doing theory, and the focus was on outcomes only; and he needed to arry the reservation prices from high to low, so that, given the sum of seller offers of quantity, he could single out a common theoretical price that "cleared the market." Thus, did he launch the neoclassical methodological break with classical economics, a

Thus, the law of demand is originally understood as a regularity holding on the aggregate of demanders; not as a direct property of individual demand behavior. This old view has been rediscovered in a more abstract version by a few mathematical economists, who, in response to the SMD problem, started to derive the law of demand by aggregation over the distribution of income or preferences (Hildenbrand, 1983; Grandmont, 1987).

**The Classical View on the Market Mechanism**

All the classical economists followed Adam Smith's explanation of the price mechanism in Chapter 7 of Book 1 of the *Wealth of Nations* ([1776] 1904), which, and this is crucial, is a unified theory of price, as opposed to the neoclassical proliferation of price theories based on the number of sellers in a market (monopoly, duopoly, oligopoly, …, "perfect competition"). These stem from the innovation of Cournot that various market experiments call into question (Smith and Williams, 1990), as does new theory inspired by experiments (Anufriev, et al. 2011).[7]

---

tradition that carries down to today, and that left behind the idea of endogenous price discovery by collective interaction, and, with it, the invisible hand metaphor.

[7] Rather than objects of stand-alone modelling, monopoly, duopoly and oligopoly are market outcomes, logically inseparable from the buyer-seller contestable context from which they arise. Monopoly is ubiquitous as an adaptive consequence of local or national demand being insufficient to support more than a single enterprise, a naturally occurring outcome in classical theory. Every new product births as a monopoly, most die, for some many entrants follow, for a few there is attrition toward dominance by from 1-4 firms.

Price, in the classical view, evolves through the competition of buyers and sellers in a way that is familiar but incompatible with "price taking" or the "law" of one price. If the amount brought to market is below what buyers are willing to take at the supply price, then a competition starts on the demand side. Some buyers are *willing to pay* a higher price, and the market price rises as these buyers outbid one another: price may continue to increase "more or less above the natural price (seller cost), according as either the greatness of the deficiency, or…the eagerness of the competition" ([1776], 1904, Vol. 2, 58). If the quantity brought to market exceeds what buyers are willing to buy at the supply price, sellers are *willing to accept* a lower price. As they undersell one another, the market price "will sink more or less below the natural price (seller cost), according as the greatness of the excess" stimulates seller competition, or how important it is that they "get immediately rid of the commodity." ([1776] 1904, Vol 2., 59) Simple as it may look at first sight, this old theory of the price mechanism is in fact a deep and rigorous one. Moreover, from where the classical economists left it, we need only a few more steps of reasoning to reach a general characterization of the market mechanism, and to pin it down to a fundamental principle.

**The Essence of the Classical Market Mechanism**

The fundamental function of a market, is informational, as Hayek famously emphasized (1937, 1945). The market mechanism, in essence, consists of revealing in the best possible way the market participants' valuations of a good. If we dig deeper into the classical competitive process, we see that the market price, through competition, evolves to reveal better and better the individual valuations of the good. This can be understood intuitively as follows: a buyer, by outbidding rival buyers, brings the market price closer to his/her valuation of the good, though the intention was contrary, to buy the cheapest possible. Symmetrically, by underselling rival producers, a firm is bringing the market price closer to its cost of production, while its intention was on the contrary to sell dearest. Each entity faces discipline by the opportunity-cost terms of others. Thus, the market exhibits a collective rationality that is no part of the intentions of the individual participants.

Formally, it follows that the overall distance between the market price and the individual valuations (which is simply an integral of excess supply) is minimized—an unintended consequence of the competition of buyers and sellers (Inoua and Smith (2019). This, be it insisted, is an emergent optimization of the market considered as a whole, since none of the individual participants has either the sophistication nor the whole information needed, nor even any direct interest in this collective value revelation.[8] In the jargon of statisticians, we would say that the market participants are unthinkingly engaged in identifying a sophisticated robust statistic: the market price converges to what we call the *center of value*, which generalizes the traditional notion of competitive equilibrium (or market clearing) and which mathematicians call generically a Fréchet median. Interestingly, this characterization of the market mechanism has been suggested from the start in experimental economics, but under the name of the "excess rent hypothesis" and interpreted as the minimization of the potential cost (or "virtual rent") that it would take to bring about a competitive equilibrium (Smith, 1962, 126-132). Here we correct and fulfill that fledgling struggle, referring to it as *The Principle of Maximum Information* (PMI).

---

[8] Comparisons with and without complete information demonstrate that individuals have no effective means of utilizing revealed complete information; complete information is neither necessary nor sufficient for equilibrium convergence (Smith, 1982, proposition 6)

The principle strictly applies only to non-durable goods and services markets without re-trade, and therefore non-speculative markets. The information maximally reflected in the market price is the consumers' use-values (weighted by the demands) and the producers' costs (weighted by the supplies). Otherwise, we can have a conflict between use-value, and resale value. In a financial market of investors solely motivated by fundamentals, the information maximally reflected in price is the investors' estimations of the asset's intrinsic value; so, provided that investors' errors are not strongly correlated, the market price converges to the intrinsic value of the asset, in a way that is robust to outliers. Thus, does the PMI offer a natural foundation for the narrowly specified "efficient market hypothesis."

In a speculative market (for a re-tradable durable good or for a financial asset), however, the relevant information are the traders' expectations of future price changes. In practice, expected prices are extrapolative (for example, based on past price changes that are trend-following), thus creating a destabilizing feedback loop[9] and a bubble that can keep on growing as long as it is backed by some source of liquidity, such as bank credit. The dichotomy between perishable and re-tradable goods and its relevance for macroeconomics have also been established experimentally (Smith, Suchanek, and Williams, 1988; Porter and Smith, 1994, 2003; Dickhaut, Lin, Porter, and Smith, 2012; Palan, 2013; Gjerstad and Smith, 2014; Gjerstad, Porter, Smith, and Winn, 2015; Smith and Inoua (2019). This explanation of economic crises and depressions as debt-fueled speculative bubbles coming to an end also goes back to the classical economists. Thus did Adam Smith famously explain the South Sea Company Bubble ([1776] 1904, Vol 2., 233), and also did J. B. Say in his important but usually overlooked *Cours complet* ([1828] 2010, part III, ch. XIX) and J. S. Mill ([1848] 1909), bk. III, ch. XII) explain more systematically economic crises. [10]

---

[9] This view echoes the complex-systems (agent-based) approach to financial markets, which treats the extreme (i.e. non-Gaussian, power-law) randomness of financial prices as an endogenous dynamics of financial markets

**Adam Smith on Beneficence and Justice: Significance for Wealth of Nations**

Adam Smith's ([1759] 1976) first book was generally ignored, if not trivialized, by twentieth century economists (Smith and Wilson, 2019, 2-4). A consequence is that his contributions to social exchange theory are usually not part of how we think about Smith's widely acclaimed

---

(Bouchaud, 2011). One can also show that the extreme (power-law) fluctuations of financial prices follow generically by definition of speculation and extrapolative (trend-following) expectations (Inoua, 2016a, 2016b).

[10] The classical view on economic instability is also rediscovered in a more sophisticated version notably by Irving Fisher, Minsky, and Kindleberger (I. Fisher, 1933; Minsky, 1992; Kindleberger and Aliber, 2011; Keen, 2013).

contributions to economics, nor can we appreciate the scope of his deep insights into the meaning and interpretation of his observations of the social world around him.

For Smith the two great pillars of society are beneficence and justice. Beneficence refers to a pattern of conduct wherein intentional acts of kindness toward others invoke an urge to reward the actor because of the feelings of gratitude the action excites in others (Smith ([1759] 1976), 78-9). It is because such actions are most likely to be directed to those who have been beneficent toward us, that it follows that kindness begets kindness. Thus, Smith derives reciprocity norms from its roots in beneficence ([1759] 1976, 225).

Beneficence is the foundation for reciprocal social exchange, which is self-enforcing in permanent communities where there is always a tomorrow to support the return of good offices for those conveyed yesterday. The implications for Smith's *Wealth of Nations* are that beneficence supports trade wherein the reciprocal benefit-reward calculus occurs simultaneously in trading partners and accounts for "the propensity to truck, barter and exchange one thing for another" ([1776] 1904, 15). Moreover, dependence on mutual trust and trustworthiness is reduced if there are mechanisms for third party enforcement of contracts, allowing the benefits of reciprocity to extend to strangers.

Such mechanisms are contained in the obverse of beneficence, Smith's second pillar of society. Justice refers to a pattern of conduct in which intentionally hurtful actions toward others provoke punishment in response because of the resentment the action excites in others ([1759] 1976, 78-9). Between the social forces of beneficence and justice, however, "we feel ourselves to be under a stricter obligation to act according to justice, than agreeably to friendship, charity or generosity…" In this, we feel supported "with the utmost propriety, and the approbation of all mankind…" ([1759] 1976, 80)

Why, since Smith's proposition is concerned with our resentful response to hurtful actions of injustice, does he call it "justice?" Because for Smith, justice is defined negatively—the absence of injustice. A world of justice is one that has enumerated specifically prohibited actions, assigned appropriate punishments—neither too large or too small, but fit for the infraction—and then allows the infinitely large set of remaining actions to be pursued freely without external constraint.[11] Thus, death excites the greatest resentment, and murder is the greatest crime. To lose our possessions is a greater evil than to be disappointed only in our expectations of gain. This is why breach of property—theft and robbery—that take that which we are possessed of, constitute greater crimes than violation of contract[12] ([1759] 1976, 84). Civil governments that incorporate and follow the "rule of law" simply codify these norms as 'shalt nots' of justice based on cultural evolutionary experience.

---

[11] Justice as liberating in this sense is contained in famous passages in Smith ([1776] 1904, 168, 184) but the foundation is in the cited pages in his first book.

[12] The greater penalty for breach of property than of contract originates in our experience of loss, which is much more intense than our experience of gain; Smith derives this proposition from a more fundamental psychological asymmetry between our joy and our sorrow. "We suffer more…when we fall from a better to a worse situation, than we ever enjoy when we rise from a worse to a better. Security, therefore, is the first and the principal object of prudence." ([1759] 1976, 213 and 45)

Thus, Smith's *Theory of Moral* Sentiments develops the property right foundations necessary for the *Wealth of Nations* to which he adds the sufficient condition—his axiom of discovery—the "propensity to truck barter and exchange" ([1776] 1905, 15). However, the latter originates in beneficence, reciprocal social exchange, and is supported, within the civil order of government, by justice in the form of third party enforcement of contracts as well as its enforcement of property.

Adam Smith's two works combine to form a comprehensive unified examination of the nature and causes of human betterment embedded in our propensity for social and economic exchange.